\documentstyle[prd,aps]{revtex}

\newcommand{\bea}{\begin{eqnarray}}
\newcommand{\eea}{\end{eqnarray}}

\begin{document}
\draft
\twocolumn[\hsize\textwidth\columnwidth\hsize\csname
@twocolumnfalse\endcsname

\title{Identification of perturbation modes and
       controversies in ekpyrotic perturbations}
\author{Jai-chan Hwang${}^{(a)}$ and Hyerim Noh${}^{(b)}$ \\
        ${}^{(a)}$ {\sl Department of Astronomy and Atmospheric Sciences,
                        Kyungpook National University, Taegu, Korea} \\
        ${}^{(b)}$ {\sl Korea Astronomy Observatory, Taejon, Korea}
        }
\date{\today}
\maketitle

\begin{abstract}

If the linear perturbation theory is valid through the bounce, 
the surviving fluctuations from the ekpyrotic scenario (cyclic one as well)
should have very blue spectra with suppressed amplitude for the 
scalar-type structure.
We derive the {\it same} (and consistent) result using the
curvature perturbation in the uniform-field (comoving) gauge 
and in the zero-shear gauge.
Previously, Khoury {\it et al.} interpreted results from the latter gauge 
condition incorrectly and claimed the scale-invariant spectrum,
thus generating controversy in the literature.
We also correct similar errors in the literature based on 
wrong mode identification and joining condition.
No joining condition is needed for the derivation.

\end{abstract}


\vskip2pc]

\section{Introduction}
                                    \label{sec:Introduction}

The issue of scalar-type structure generated in the
recently proposed ekpyrotic scenario is shrouded with controversies
by two opposing camps \cite{KOST,Durrer,Peter-Pinto-Neto-2002} and 
\cite{Lyth-1,BF-2001,Hwang-2002-Ekpyrotic,Lyth-2,Hwang-Noh-2002-Bounce,Martin-etal-2001,Tsujikawa}; for an introduction to the scenario, see \cite{Rasanen}.
The main point of \cite{KOST} is that the
dominating solution viewed in the zero-shear hypersurface (gauge)
in the collapsing phase happens to show a scale-invariant spectrum.
However, this mode was identified in \cite{BF-2001,Hwang-2002-Ekpyrotic}
as a transient mode in the subsequent expanding phase, thus uninteresting.
In this work we wish to add some additional points to 
\cite{Hwang-2002-Ekpyrotic}.
We will show that the same blue spectrum is generated even in the 
zero-shear gauge by identifying the mode relevant in the later expanding phase.
Apparently, the same final observable spectrum should be derived independently
of the gauge conditions used, and our result confirms it.
We also point out that the possible scale-invariant spectra and others
argued in \cite{Durrer,Peter-Pinto-Neto-2002,FB-2002} are errored 
by identifying wrong modes (often based on {\it ad hoc} joining conditions)
which are transient in the expanding phase, thus irrelevant.

Before we embark on studying the evolution of structures through
bounce using the linear perturbation theory, we would like to 
state clearly the provisions we need.
In \cite{Lyth-2} Lyth has clearly shown that the linear perturbation theory 
{\it breaksdown inevitably} as the model approaches the singularity in a 
singular bounce (if such a bounce is possible at all, \cite{KOSST}), 
see also \S VI of \cite{Hwang-Noh-2002-Bounce}.
Somehow, this strong conclusion is unduely ignored by many authors 
\cite{KOST,Durrer}.
If the bounce is singular we cannot rely on the linear 
perturbation theory.
Thus, in the ekpyrotic scenario and other bouncing models considered 
in this paper we will explicitly {\it assume} that the bounce occurs 
before the linear theory breaks down, and this {\it requires}
the bounce to be smooth and nonsingular.
Although the authors of \cite{KOST} claimed that the bounce
in the ekpyrotic scenario to be singular,
assuming a nonsingular bounce in such a scenario is legitimate, 
particularly if we consider the currently unknown physics near the bounce.
Since the consequence of the singular bounce is clearly resolved
in \cite{Lyth-2} ({\it i.e.}, the linear theory fails!) 
investigating the remaining window with nonsingular bounce
would be important to clearly resolve the remaining issue.

In a single component fluid or field, the scalar-type perturbation
is described by a second-order differential equation with two solutions (modes).
In the large-scale limit (to be defined later) we can often derive
a general asymptotic solution with two modes, see eq. (\ref{LS-solutions}).
In an expanding phase we can identify clearly which ones are relatively
growing ($C$-mode) and decaying ($d$-mode).
If the initial condition is imposed at some early expanding epoch 
the decaying mode is transient in time, and naturally we are only 
interested in the relatively growing mode.
If we introduce a collapsing phase before the early big-bang phase,
however, the conventional growing and decaying classification
can be often reversed.
Still, if the large-scale conditions are met (and, of course, 
{\it if} the linear theory as well as the classical gravity are intact), 
the general solutions
in eq. (\ref{LS-solutions}) remain valid throughout the transition.
Thus, in our observational perspective situated in expanding phase
we are interested in the initial condition imposed on the $C$-mode,
even if it {\it was} subdominating (relatively decaying) 
compared with the other mode when the initial condition was imposed.
Although we made this point clear in \cite{Hwang-2002-Ekpyrotic},
in this work we will reinforce it by deriving concretely the
$C$-mode initial conditions coming from the quantum vacuum fluctuations
in the two gauge conditions used previously.
In this way, we hope we could clear some of the controversies 
concerning ekpyrotic scenario and others in the literature.
\S \ref{sec:Equations} and \ref{sec:Power-law} are reviews.
\S \ref{sec:Identification} contains our main results with 
consequences analysed in \S \ref{sec:Discussions}.
We set $c \equiv 1 \equiv \hbar$.

\section{Basic equations and general large-scale solutions}
                                    \label{sec:Equations}

We consider the scalar-type perturbation in a flat Friedmann 
world model supported by a minimally coupled scalar field.
Our metric convention follows Bardeen's in \cite{Bardeen-1988}
\bea
   ds^2
   &=& - a^2 (1 + 2 \alpha) d\eta^2
       - 2 a^2 \beta_{,\alpha} d \eta d x^\alpha
   \nonumber \\
   & & + a^2 \left[ g^{(3)}_{\alpha\beta} ( 1 + 2 \varphi )
       + 2 \gamma_{,\alpha|\beta} \right] dx^\alpha dx^\beta,
\eea
and $\chi \equiv a (\beta + a \dot \gamma)$;
an overdot and a prime indicate time derivatives based on $t$
and $\eta$, respectively, with $dt \equiv a d \eta$.
The background is described by
\bea
   & & H^2 = {8 \pi G \over 3} \left( {1 \over 2} \dot \phi^2 + V \right), 
       \quad
       \ddot \phi + 3 H \dot \phi + V_{,\phi} = 0,
\eea
where $H \equiv {\dot a \over a}$.
The basic perturbation equations are \cite{Field-Shepley,Lukash-Mukhanov}:
\bea
   & & u = - {4 \pi G z \over k^2} \left({v \over z} \right)^\prime, \quad
       v = {1 \over 4 \pi G z} \left( z u \right)^\prime,
   \label{u-v-relations} \\
   & & v^{\prime\prime} + \left[ k^2 - {z^{\prime\prime} \over z}
       \right] v = 0, \quad
       u^{\prime\prime} + \left[ k^2 - {(1/z)^{\prime\prime} \over 1/z}
       \right] u = 0,
   \label{u-v-equations} 
\eea
where $z \equiv {a \dot \phi / H}$, and
\bea
   & & v \equiv a \delta \phi_\varphi, \quad
       \varphi_{\delta \phi} \equiv \varphi - (H / \dot \phi) \delta \phi
       \equiv - (H / \dot \phi) \delta \phi_\varphi,
   \nonumber \\
   & & u \equiv - \varphi_\chi/\dot \phi, \quad
       \varphi_\chi \equiv \varphi - H \chi.
   \label{GI-def}
\eea
$\varphi_{\delta \phi}$ and $\varphi_\chi$ are gauge-invariant combinations
which are equivalent to the curvature perturbation ($\varphi$)
in the uniform-field gauge ($\delta \phi \equiv 0$, equivalently 
the comoving gauge) and in the zero-shear gauge ($\chi \equiv 0$), 
respectively \cite{Bardeen-1988}.
The perturbed action was derived in \cite{Lukash-Mukhanov}
\bea
   & & \delta^2 S = {1 \over 2} \int
       \left( v^{\prime 2} - v^{|\alpha} v_{,\alpha}
       + {z^{\prime\prime} \over z} v^2 \right) d^3 x d \eta.
   \label{action}
\eea

In the large-scale limit, meaning for negligible $k^2$ terms, 
eq. (\ref{u-v-equations}) has general solutions 
\cite{Field-Shepley,Lukash-Mukhanov,Hwang-Noh-2002-Bounce}
\bea
   \varphi_{\delta \phi} (k, \eta) 
   &=& C (k) - d(k) {k^2 \over 4 \pi G} \int^\eta {d \eta \over z^2}, 
   \nonumber \\
   \varphi_\chi (k, \eta)
   &=& 4 \pi G C (k) {H \over a} \int^\eta z^2 d \eta + {H \over a} d (k).
   \label{LS-solutions}
\eea
To the higher-order in the large-scale expansion
each of the four solutions have 
$[1 + \sum_{n=1,2,3, \dots} \tilde c_n (k |\eta|)^{2n}]$ factor
with $\tilde c_n$ differing for the four cases.
We {\it emphasize} the general nature of these solutions
in the large-scale limit.
These are exact solutions of the spatial curvature perturbation
($\varphi$) in the respective hypersurfaces (gauges)
valid as long as the $k^2$ terms in eq. (\ref{u-v-equations}) are negligible;
thus valid for general (time-varying) potential $V (\phi)$.
Similar general solutions exist for the fluid situation 
for general (time-varying) equation of state $P (\mu)$,
and even for the generalized gravity theories \cite{LS-sols}.

Schwarz has pointed out that as a smooth bounce has to violate 
the weak energy condition if space-time is flat, $z^2 \propto \mu + P$ 
will have (at least two) zeros; 
this means that $1/z$ will be ill defined and the higher order 
corrections will have singular coefficients, thus the long wavelength 
expansion becomes inconsistent, \cite{Schwarz}.
To achieve a bounce, in \S V.C of \cite{Hwang-Noh-2002-Bounce} we have used an 
additional presence of an exotic matter $X$ with negative energy density.
Thus, such an $X$-matter cannot dominate even during the bouncing phase.
In \cite{Hwang-Noh-2002-Bounce} we have shown that if we concentrate on the
evolution of {\it curvature perturbation}, assuming near adiabatic
initial condition in the collapsing phase, eq. (\ref{u-v-equations}) for $u$, 
thus our solution for $\varphi_\chi$ in eq. (\ref{LS-solutions}) as well,
{\it remains valid}.
In this context, $z^2$ goes through
vanishing points at least twice in the bouncing phase, and
indeed, in that case the next order large-scale
expansion includes $\int (1/z^2) d\eta$-order terms which are ill defined;
one such term already appears in the $d$-mode of $\varphi_{\delta \phi}$
in eq. (\ref{LS-solutions}) which is exactly the next order contribution.

As parts of the series solutions in eq. (\ref{LS-solutions})
are ill defined for $z = 0$, in such a case we
should {\it go back} to our original equations (forms before we combine
to make a second-order equation).
One such original equation is conveniently available in the second equation of
eq. (\ref{u-v-relations}) which shows that, for $z = 0$ we have
$(\varphi_\chi a/H)^\prime = 0$.
Thus, for $z=0$ we have an exact solution: $\varphi_\chi \propto H/a$.
Notice that our eq. (\ref{LS-solutions}) {\it includes} the 
above solution as a case!
Therefore, we conclude that throughout the bounce (including $z=0$ points), 
our leading order aymptotic solution
for $\varphi_\chi$ in eq. (\ref{LS-solutions}) remains valid.
Thus, the ill defined higher order corrections in the series expansion 
do not cause any practical problem in the perturbations.

\section{Power-law expansion}
                                    \label{sec:Power-law}

A field with an exponential potential supports power-law expansion/contraction
of the scale factor \cite{Lucchin-Matarrese-1985}
\bea
   & & a \propto |t|^p \propto |\eta|^{p/(1-p)}, \quad
       V = - {p ( 1 - 3 p) \over 8 \pi G} e^{-\sqrt{16 \pi G /p} \phi},
   \nonumber \\
   & & H/\dot \phi = \sqrt{4 \pi G p}.
   \label{power-law}
\eea
In the power-law case eq. (\ref{u-v-equations}) leads to
Bessel equations for $v$ and $u$ with different orders. 
Using the quantization based on the action formulation
in eq. (\ref{action}), we have the exact mode function solutions
($p \neq 1$) \cite{Lyth-Stewart,Hwang-2002-Ekpyrotic}
\bea
   \varphi_{\delta \phi k}  (\eta)
   &=& \left| {H \over \dot \phi} \right| {\sqrt{ \pi |\eta|} \over 2 a}
       \Big[ c_1 (k) H_{\nu_v}^{(1)} (x)
       + c_2 (k) H_{\nu_v}^{(2)} (x) \Big],
   \nonumber \\
   \varphi_{\chi k} (\eta)
   &=& { |H| \sqrt{\pi^2 G |\eta|} \over k \sqrt{p} }
       \Big[ c_1 (k) H_{\nu_u}^{(1)} (x)
       + c_2 (k) H_{\nu_u}^{(2)} (x) \Big],
   \nonumber \\
   & & \nu_v \equiv {3 p - 1 \over 2 (p - 1)}, \quad
       \nu_u \equiv {p + 1 \over 2 (p - 1)},
   \label{power-law-solution}
\eea
where $x \equiv k |\eta|$. 
The quantization condition implies $|c_2|^2 - |c_1|^2 = \pm 1$ 
depending on the sign of $\eta$, \cite{Hwang-2002-Ekpyrotic}.

\section{Mode identification}
                                    \label{sec:Identification}

The Hankel functions can be expanded as
\cite{Abramowitz-Stegun-1972}
\bea
   H_\nu^{(1,2)} (x)
   &=& \sum_{n=0}^\infty {1 \over n!} \left(- {x^2 \over 4} \right)^n
       {1 \over \sin{\nu\pi}}
       \Bigg[ \left( {x \over 2} \right)^\nu 
       { \pm i e^{\mp i \nu \pi} \over \Gamma(\nu + n + 1) }
   \nonumber \\
   & & + \left({ x \over 2} \right)^{-\nu} 
       { \mp i \over \Gamma (- \nu + n + 1) } \Bigg].
   \label{Hankel-expansion}
\eea
Notice that, in the small $x$ limit, the first (second) term in the
parenthesis dominates for $\nu <0$ ($\nu > 0$).
In eq. (\ref{LS-solutions})
the leading orders of the $C$-modes are time independent
whereas the leading orders of the $d$-modes behave as
$\varphi_{\delta \phi} \propto |\eta|^{2 \nu_v}$ and
$\varphi_{\chi} \propto |\eta|^{2 \nu_u}$.
Since
\bea
   & & \varphi_{\delta \phi k} 
       \propto |\eta|^{\nu_v} H_{\nu_v}^{(1,2)} (k |\eta|), \quad
       \varphi_{\chi k} \propto k^{-1} |\eta|^{\nu_u} 
       H_{\nu_u}^{(1,2)} (k |\eta|),
   \nonumber \\
\eea
we can easily identify the first and the second terms
in the parenthesis of eq. (\ref{Hankel-expansion})
as the $d$-mode and the $C$-mode, respectively.

The lower-bounds of integrations in eq. (\ref{LS-solutions}) 
give rise to terms which can be absorbed to the other modes.
Such an ambiguity is removed, for example, by identifying the $C$-mode 
of $\varphi_{\delta \phi}$ in expanding phase in the large-scale limit
when the time-dependent part of the $d$-mode has asymptotically decayed away.
The authors of \cite{Leach-etal} have shown that the $d$-mode does not
necessarily decay away immediately after the horizon crossing;
in some inflationary models we have the $d$-mode effect
not negligible near the horizon-crossing, and the final
result can be interpreted as an amplification of the spectrum.
Such an amplification occurs because, in expanding phase, it takes some time to
have the time-dependent part of $d$-mode be negligible.
The point is that there is no such an effect from the $d$-mode {\it while}
in the asymptotically super-horizon scale.
In our case of the bounce the relevant scale remains in the
asymptotically super-horizon scale during the bounce, thus the
solutions in eq. (\ref{LS-solutions}) are well valid, and we do not anticipate
any ambiguity arising while in the asymptotically super-horizon scale.
Accordingly, later in \S \ref{sec:Identification}, we will identify 
the $C$- and $d$-modes ignoring the contributions from lower-bounds 
of integrations of eq. (\ref{LS-solutions}).

The power spectrum and the spectral index are defined as
${\cal P}_{\varphi} = {k^3 \over 2 \pi^2} |\varphi_k|^2$
and $n_S - 1 \equiv d \ln{{\cal P}_\varphi}/ d \ln{k}$.
The spectral indices for the $C$-modes of 
$\varphi_{\delta \phi}$ and $\varphi_\chi$ can be read as
(in the following, we {\it assume} the simplest vacuum state choice)
\bea
   & & (n_S - 1)_{\varphi_{\delta \phi}, C} 
       = (n_S - 1)_{\varphi_\chi, C} = {2 \over 1 - p}.
   \label{n_S-C-mode}
\eea
Although not interesting (because it becomes transient in an expanding phase)
the spectral indices for the $d$-modes are
\bea
   & & (n_S - 1)_{\varphi_{\delta \phi}, d} = {4 - 6p \over 1 - p}, \quad
       (n_S - 1)_{\varphi_\chi, d} = - {2p \over 1 - p}.
   \label{n_S-d-mode}
\eea
Notice that the spectral indices of the $C$-mode {\it coincide}
in both gauge conditions, whereas the ones for the $d$-mode
show {\it strong gauge dependence}.
This is easily understandable from the general solutions
in eq. (\ref{LS-solutions}):
in the power-law expansion case 
we have $\varphi_{\delta \phi}/ \varphi_\chi = 1 + p$ for the $C$-mode;
the $C$-mode of $\varphi_{\delta \phi}$ remains constant even
under the changing potential whereas $\varphi_\chi$ changes it value.
Similarly, for the $d$-mode we have
$\varphi_{\delta \phi}/ \varphi_\chi = {(p-1)^2 \over 3p-1} (k |\eta|)^2$,
thus we have 
$(n_S -1)_{\varphi_\chi, d} = (n_S -1)_{\varphi_{\delta \phi}, d} - 4$.

We note again that the $d$-modes show strong gauge dependence:
the $d$-mode of $\varphi_{\delta \phi}$ shows more blue spectrum
compared with $\varphi_\chi$.
Near singularity, the $d$-mode of $\varphi_\chi$ diverges more
strongly compared with the ones in the other gauge conditions \cite{LS-sols};
see \S III of \cite{Hwang-Noh-2002-Bounce} for a summary. 
The strong divergence in the zero-shear gauge is known to be due to
the strong curvature of the hypersurface (temporal gauge condition)
\cite{Bardeen-1980}. 
According to Bardeen the behavior of $\varphi_\chi$ 
``overstates the physical strength of the singularity'', \cite{Bardeen-1980}.
Thus, even in the collapsing phase where $d$-mode is the proper growing
solution, we {\it should not} attach more meaning to the $d$-mode 
of $\varphi_\chi$ than to the one of $\varphi_{\delta \phi}$.

\section{Consequences}
                                    \label{sec:Discussions}

In an ekpyrotic scenario with $0< p \ll 1$ 
we have a very blue $n_S - 1 \simeq 2$ spectrum for the $C$-mode.
Although $n_S - 1 \simeq 0$ for the $d$-mode of $\varphi_\chi$,
we are not interested in the $d$-mode. 
Incidentally, we have $n_S - 1 \simeq 4$ for the $d$-mode of 
$\varphi_{\delta \phi}$ which {\it better} characterizes the physical 
strength of the growing perturbation during the collapsing phase
{\it than} $\varphi_\chi$.
Our point is that, although the $d$-mode is the relatively growing solution 
in the collapsing phase, our classification of the
$C$- and $d$-modes is based on the {\it general} large-scale
solutions in eq. (\ref{LS-solutions}).
The large-scale conditions used to get these solutions
are well met during the transition phase in the ekpyrotic scenario.
In \cite{Hwang-Noh-2002-Bounce} we have shown {\it analytically} that,
as long as the linear perturbation is valid,
the solutions in eq. (\ref{LS-solutions})
remain valid throughout a (smooth and nonsingular) bounce, 
thus there occurs no mixing 
for the eventual growing solution in the later expanding phase.
Thus, claiming the scale-invariant spectrum based on
the $d$-mode of $\varphi_\chi$ is {\it incorrect}; 
see the next paragraph for some technical details.
The $C$-modes of both $\varphi_{\delta \phi}$ and $\varphi_\chi$
show the same blue spectra, and
the complete spectrum of the $C$-mode and the one for the tensor-type 
perturbation can be found in \cite{Hwang-2002-Ekpyrotic}.

We would like to comment on several minor complications made in \cite{KOST}.
{}Firstly, the authors of \cite{KOST} claimed that by combining 
the scale-invariance of the $d$-mode of $\varphi_\chi$
before the bounce and the 
{\it new} joining condition {\it introduced by the authors} they can
derive a scale-invariant final spectrum. 
This implies that mixing occurs so that the $d$-mode before the
bounce sources and dominates the $C$-mode in the subsequent
expanding phase.
This {\it contradicts} with our result based on the general 
large-scale solutions in eq. (\ref{LS-solutions}).
It was shown in \cite{BF-2001,Hwang-2002-Ekpyrotic} that the well known 
joining condition based on equations of motion \cite{Hwang-Vishniac-1991} 
in fact confirms our result: {\it i.e.}, in the large-scale
limit the $C$-mode is affected only by the $C$-mode of the previous era.
The new joining condition used in \cite{KOST} is {\it ad hoc} 
and is not based on proper physical or mathematical arguments, 
see \cite{Hwang-2002-Ekpyrotic,Martin-etal-2001}.
As we have shown in this paper, and more properly in 
\cite{Hwang-Noh-2002-Bounce}, in order to trace the large-scale 
evolution of the eventual $C$-mode in expanding phase,
we can use the analytic solutions in eq. (\ref{LS-solutions}), 
thus we do not need the joining condition at all.
Secondly, in \cite{KOST} it was emphasized that before the bounce the potential 
is restored to zero so that expansion rate changes to $p \simeq {1 \over 3}$. 
As the perturbation still remains in superhorizon scale during the bounce
such a change in the field potential does not affect the already
generated perturbation spectrum.
We have emphasized that the large-scale general solution in
eq. (7) remains valid even under such a changing potential.
We are interested only in the $C$-mode and the solution shows that 
the $C$-mode is not affected by the changing potential.
Thirdly, the authors of \cite{KOST} also stressed that radiation 
is present after the bounce.
In \S V.C of \cite{Hwang-Noh-2002-Bounce} we have shown that
the evolution of adiabatic (curvature) perturbation is not affected by 
the changing background equation of state or the presence of 
multiple component while in the superhorizon scale.
Thus, the presence of radiation component after the bounce adds only
a minor complication which does {\it not} affect the curvature 
perturbation in the superhorizon scale.

In a similar context, for $p = {2 \over 3}$ 
we have $n_S - 1 = 0$ for the $d$-mode of $\varphi_{\delta \phi}$
(in this case we have $n_S - 1 = - 4$ for the $d$-mode of $\varphi_\chi$).
Identifying this as another possibility for generating
a scale-invariant spectrum attempted in \cite{FB-2002}
is {\it incorrect} for the same reason as in the ekpyrotic case;
this was pointed out in \cite{Hwang-2002-Ekpyrotic}.
{}For the $C$-mode we have $n_S -1 = 6$, thus too blue.

Another similar error was made in \cite{Peter-Pinto-Neto-2002},
now in the case of $p = {1 \over 2}$.
In this case we have $n_S - 1 = 4$ for the $C$-modes, and 
$2$ for $\varphi_{\delta \phi}$ and $-2$ for $\varphi_\chi$ for the $d$-modes. 
Based on the zero-shear gauge authors of \cite{Peter-Pinto-Neto-2002}
claimed that the generated spectrum has $n_S = -1$ for $p = {1 \over2}$
and $n_S = 1$ for $p \simeq 0$ (ekpyrotic!), both of which are the ones for
the $d$-mode of $\varphi_\chi$, thus irrelevant 
for the final surviving (observationally relevant) spectrum.
Although \cite{Peter-Pinto-Neto-2002} used a bounce model 
which differs slightly from the one used in \cite{Hwang-Noh-2002-Bounce},
as we have argued, while in the superhorizon scale the final
surviving spectrum is independent of the changing background expansion rate.
The authors of \cite{Peter-Pinto-Neto-2002} also considered a
radiation dominated era during the quantum generation stage
whereas we considered a scalar field dominated era with $p={1 \over 2}$.
It is well known that the scalar field with an exponential
potential can be effectively identified as an ideal fluid with
constant $w(\equiv P/\mu)$, thus the two systems coincide for $p = {1 \over 2}$.

Yet another similar errors were recently added in the literature, \cite{Durrer}.
The authors of \cite{Durrer} argued that one cannot ignore the 
entropy generation near the bounce of the ekpyrotic scenario;
if the bounce is singular, we already have stated that
the problem cannot be handled using the linear theory.
Based on this argument the authors claimed that the conventional
joining conditions should be changed.
Unless we use the proper joining conditions {\it derived} in 
\cite{Hwang-Vishniac-1991}, we can show that the growing (and dominating) 
$d$-mode in the collapsing phase can easily dominate and source the $C$-mode
in the subsequent expanding phase while in the large-scale.
In this way, the authors claimed $n_S = 1$ spectrum for the
ekpyrotic scenario which comes from the $d$-mode of $\varphi_\chi$.
However, we can see that the entropy generation anticipated
near bounce would {\it not} affect the superhorizon evolution of 
(the $C$-mode) perturbation.
The joining conditions known in the literature 
give the same result as our present one based on analytic solutions,
\cite{BF-2001,Hwang-2002-Ekpyrotic}.
Perhaps the entropy generation would be important for the background 
evolution so that we could achieve a smooth and nonsingular bounce as we have 
investigated using toy models in \cite{Hwang-Noh-2002-Bounce}.

The authors of \cite{Durrer} also have claimed, 
that even for the pre-big bang scenario the final spectrum should pick up the
$d$-mode of $\varphi_\chi$ generated in the collapsing phase.
{}For the pre-big bang scenario based on a conformally transformed
Einstein frame we have eq. (\ref{power-law}) with $p = {1 \over 3}$;
thus we have a vanishing potential.
In such a case we have
$n_S - 1 = 3$ for the $C$-mode, and
$3$ for $\varphi_{\delta \phi}$ and $-1$ for $\varphi_\chi$ for the $d$-modes.
Based on the same logic as their ekpyrotic case, 
the authors claimed that the final spectrum should be
$n_S = 0$ which is the one for $d$-mode of $\varphi_\chi$.
We already have explained what is wrong with such analysis and result.
The correct $n_S = 4$ spectrum in Einstein frame was derived in
\cite{Brustein-etal-1995}.
In the original frame based on the low-energy effective action of string theory 
the pre-big bang scenario 
shows a pole-like inflation with $a \propto |t-t_0|^{-1/\sqrt{3}}$.
The perturbation spectrum was derived by us
in the original frame with $n_S - 1 = 3$, \cite{Hwang-1998-PBB}.
We also have shown that $\varphi_{\delta \phi}$
is conformally invariant \cite{Hwang-1997-CT}.
Thus, it is natural for the final spectra from the two frames 
(despite their very different descriptions of the background evolutions)
to coincide. 
We note that in the orignal frame the pre-big bang
scenario does {\it not} involve a contracting phase, and is 
just another inflation. 
In such a case, as in the case of ordinary inflation, the calculation 
does not require any joining condition.

If the linear perturbation theory is valid throughout,
and the bounce is smooth and non-singular (see \cite{Hwang-Noh-2002-Bounce}
for several examples) we could rely on solutions in eq. (\ref{LS-solutions})
as long as the large-scale conditions are met.
In such a case, as emphasized in \cite{Hwang-2002-Ekpyrotic},
we do not need to use joining condition which, if we use the ones 
derived properly, also gives the same result 
\cite{BF-2001,Hwang-2002-Ekpyrotic,Hwang-Noh-2002-Bounce,Martin-etal-2001}.
Several possibilities to have (smooth and non-singular) bounce models
were studied in \cite{Hwang-Noh-2002-Bounce}.
In \cite{Hwang-Noh-2002-Bounce}, using a toy bounce model based on 
an exotic matter with a negative energy density we have shown analytically
that the pre- and post-bounce results of the $C$-modes 
of $\varphi_{\delta \phi}$ 
and $\varphi_\chi$ show the same behaviors as the ones we studied
in this work (which ignores the precise physics of the bounce), 
{\it independently} of the presence of the exotic matter 
(and the bounce itself) introduced 
to connect the collapsing and the expanding phases,
see \S V.C in \cite{Hwang-Noh-2002-Bounce}.

Equation (\ref{n_S-C-mode}) shows that the only way to get
a $n_S - 1 \simeq 0$ spectrum from the power-law expansion
based on an exponential potential is to have $p \gg 1$
which is the ordinary power-law expansion {\it or} a 
damped collapsing phase.
As pointed out in \cite{Hwang-2002-Ekpyrotic}, in the latter
case as the model approaches the bouncing phase the comoving scales
shrink faster than the Hubble (dynamical) horizon. 
Thus, the large-scale condition can be violated near the bounce, 
and we cannot simply trace the perturbation through the bounce,
see \cite{Hwang-Noh-2002-Bounce}.
Therefore, the only remaining possibility to get an observationally
viable spectrum is the former case which is just 
a well known version of inflation.

\subsection*{Acknowledgments}

We thank Robert Brandenberger, Patrick Peter, Nelson Pinto-Neto,
and Dominik Schwarz for useful correspondences.
This work was supported by Korea Research Foundation grants 
(KRF-2001-041-D00269).


\end{document}